\documentclass[11pt]{JHEP3}

\usepackage{epsfig,multicol,multirow,bbm}

\newcommand{\gev}{{\hbox{GeV}}}

\newcommand{\mgev}{{\hbox{GeV}/c^2}}
\newcommand{\m}{{\hbox{m}}}
\newcommand{\cm}{{\hbox{cm}}}
\newcommand{\g}{{\hbox{g}}}

\newcommand{\mrad}{{\hbox{mrad}}}
\newcommand{\s}{{\hbox{s}}}
\newcommand{\lumi}{\cm^{-2} \s^{-1}}

  \title{The estimation of neutrino fluxes produced by proton-proton collisions at $\sqrt{s}=14$~TeV of the LHC }
  \author{HyangKyu~Park, \\
  Center for High Energy Physics, Kyungpook National University, Daegu, 702-701, Korea\\
  E-mail: \email{hkpark@knu.ac.kr}}

\abstract{
Intense and collimated neutrino beams are produced by charm and beauty particle decays from proton-proton collisions at the LHC. 
A neutrino experiment would be run parasitically without interrupting the LHC physics program  during the collider run. 
We estimate the neutrino fluxes from proton-proton collisions at $\sqrt{s}=14$~TeV of 
the LHC with the designed luminosity, $10^{34} \lumi$. 
By mounting about 200 tons  of fiducial volume of a neutrino detector at 300 $\m$ away from the interaction point, 
about 150,000 of charged current neutrino events per year can be observable.
}
  

\begin{document}

\section{Introduction}
The main objectives of the LHC (Large Hadron Collider) at CERN are to search for an origin of electroweak symmetry breaking and to probe a new physics in the TeV scale. When the LHC is fully operational, it will have proton-proton collisions at center-of-mass energy of 14 TeV  
with a design luminosity of $10^{34} \lumi$. In a long term, the LHC is planning to upgrade to  the super-LHC (sLHC), which is under evaluation.
The sLHC will deliver 10 times more luminosity, $10^{35} \lumi$, than the LHC.

In such a huge luminosity at the LHC, there will be  intense and collimated neutrino beams
from particle decays produced by the collisions~\cite{rujula1}.  
The neutrino and anti-neutrino production rates at the interaction point  would be equal in forward and backward directions.
By mounting a neutrino detector at $\sim$300 $\m$ away from the interaction point, 
one can perform a neutrino experiment parasitically without interrupting the LHC physics program  during the collider run.
And also, this detector would provide an unique opportunity for a long-lived metastable particle searche, for example,  
the scalar tau lepton~\cite{feng,hamaguchi} in 
Supersymmetric extension of gravity, as well as neutrino physics.  
Another advantage of this experiment is that
the neutrino fluxes and momentum spectrum could be well understood since the collider detectors, CMS and ATLAS, at LHC  can be used for
monitoring production rates of major neutrino sources, Pions, Kaons, heavy mesons and {\it etc.}.   

In Ref~\cite{rujula2}, the neutrino fluxes from decays of charm and beauty mesons produced at 
the collision point of LHC at center-of-mass energy of 16 TeV with an luminosity of $2 \times 10^{34} \lumi$
were estimated.  The Quark Gluon String Model~\cite{kaidalov} was used for
the estimation of the inclusive production rates for charm and beauty mesons, and  Hagedorn's  thermodynamic model~\cite{hagedorn} was adopted 
for describing the transverse momentum distributions for the produced heavy mesons. 
About 1,000 of charged current $\nu_{\tau}$ and $\bar{\nu}_{\tau}$ interactions per year were expected with 2.4 (62.4) tons of detector locating at 100 (500) $\m$ from the interaction point.

In this paper, we use the PYTHIA event generator~\cite{pythia} with the parameter set~\cite{proq20} tuned by experimental data from Tevatron and LEP
to estimate neutrino fluxes from the particle decays produced by proton-proton collisions at   $\sqrt{s}=14$~TeV 
with the luminosity of $10^{34} \lumi$. 
Once  a parameter set for the PYTHIA is tuned using experimental data produced by 
the LHC at $\sqrt{s}=14$~TeV, we will have better estimation for
neutrino fluxes.

\section{Estimation of Neutrino Fluxes}
Since main sources of neutrinos from pp collisions at the LHC are Pions, Kaons, charm and beauty particles, 
the yields and transverse and longitudinal momentum spectra of those particles should be well understood for the estimation of neutrino fluxes.  
We use the PYTHIA program with the tuned parameter set, which is
widely used and intensively tested with real data for the LHC experiments at $\sqrt{s}=7$~TeV. 
Although the yields for Kaons and beauty mesons are not well agreed with the
PYTHIA~\cite{strange,beauty1,beauty2}, the shape for transverse momentum spectra of those particles are generally good.
Recent measurement for $\sigma(pp \to b \bar{b}X)$ at $\sqrt{s}=7$~TeV is found to be $(284 \pm 20 \pm 49)~\mu b$~\cite{bbxsec}, 
 and PYTHIA estimation is $226.7~\mu b$, which is about 20~\% lower. This suggests that we may underestimate the neutrino flux by about 20~\% level.
    
\begin{table}
 \centering
 \begin{tabular}{|c|c|c|c|c|c|c|}
\cline{1-7}
  Neutrino              & \multicolumn{3}{|c|}{Average neutrino-beam energy ($\gev$)}  & 
                               \multicolumn{3}{|c|}{~~~~~~ No. of neutrino per year ($10^{12}$)} \\ \cline{2-7} 
  Beam                                    & ~~Light & ~~Heavy & All & ~~Light & ~~Heavy & All  \\ \cline{1-7}
  $\nu_e$     &  102.9 & 777.4 & 747.1 & 0.14  & 2.92 & 3.06 \\ \cline{1-7}
  $\nu_\mu$ &  81.4  & 911.7 & 540.5  & 4.35   &  5.38 & 9.73 \\ \cline{1-7}
  $\nu_\tau$ &   - & 206.5 & 206.5 & - & 0.073 & 0.073 \\ \cline{1-7}
\end{tabular}
 \caption{ Summary of the average neutrino energy and the number of neutrinos including anti-neutrino per year 
for neutrino within 3 $\mrad$ of cone
in either forward or backward direction with $10^{34} \lumi$ of the nominal luminosity. 
The columns for Light, Heavy and All show 
the contributions from light, heavy, and light and heavy particle decays, respectively.}
\label{tab:energy}
\end{table}

 \FIGURE{
               \epsfig{file=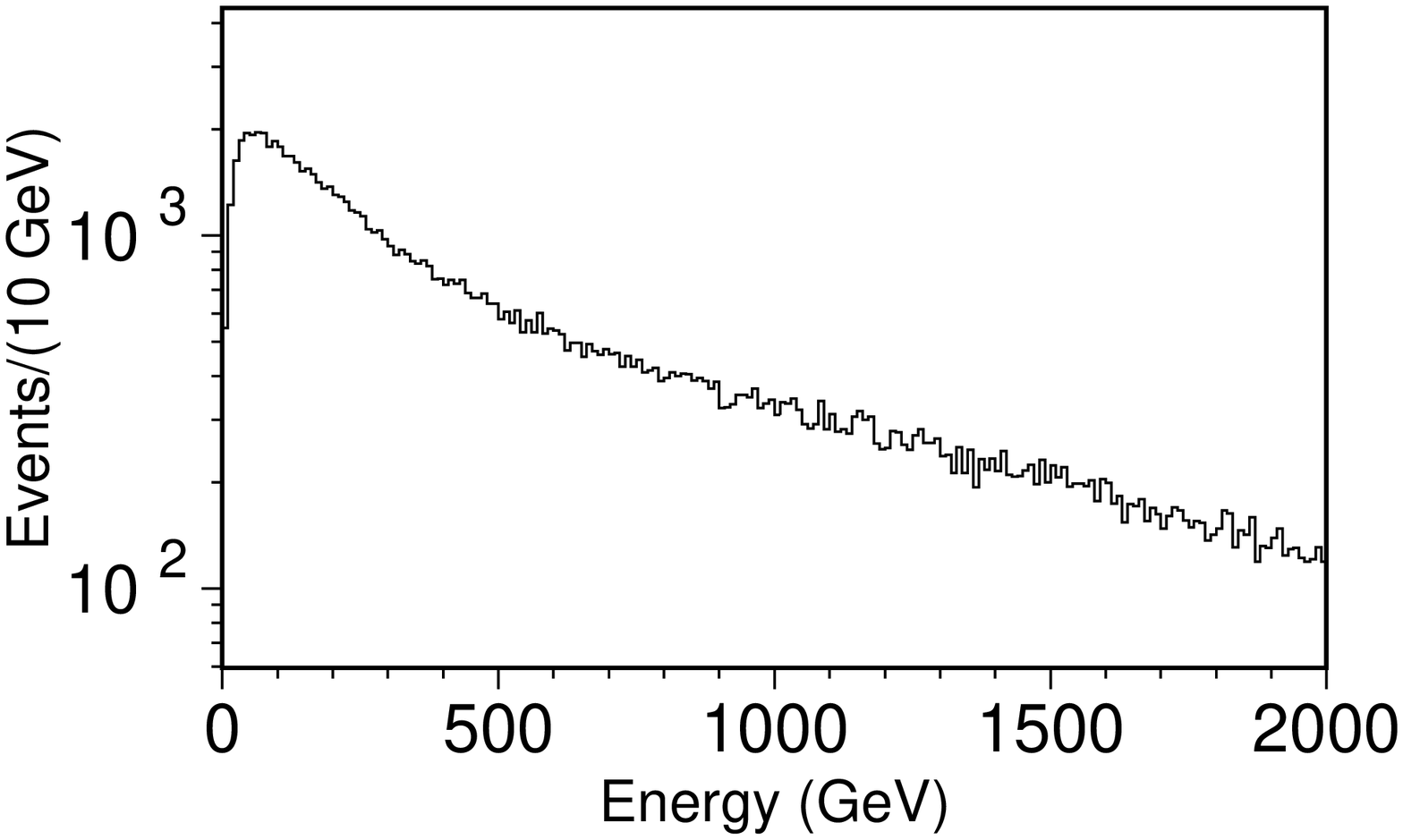,width=10cm}
              \epsfig{file=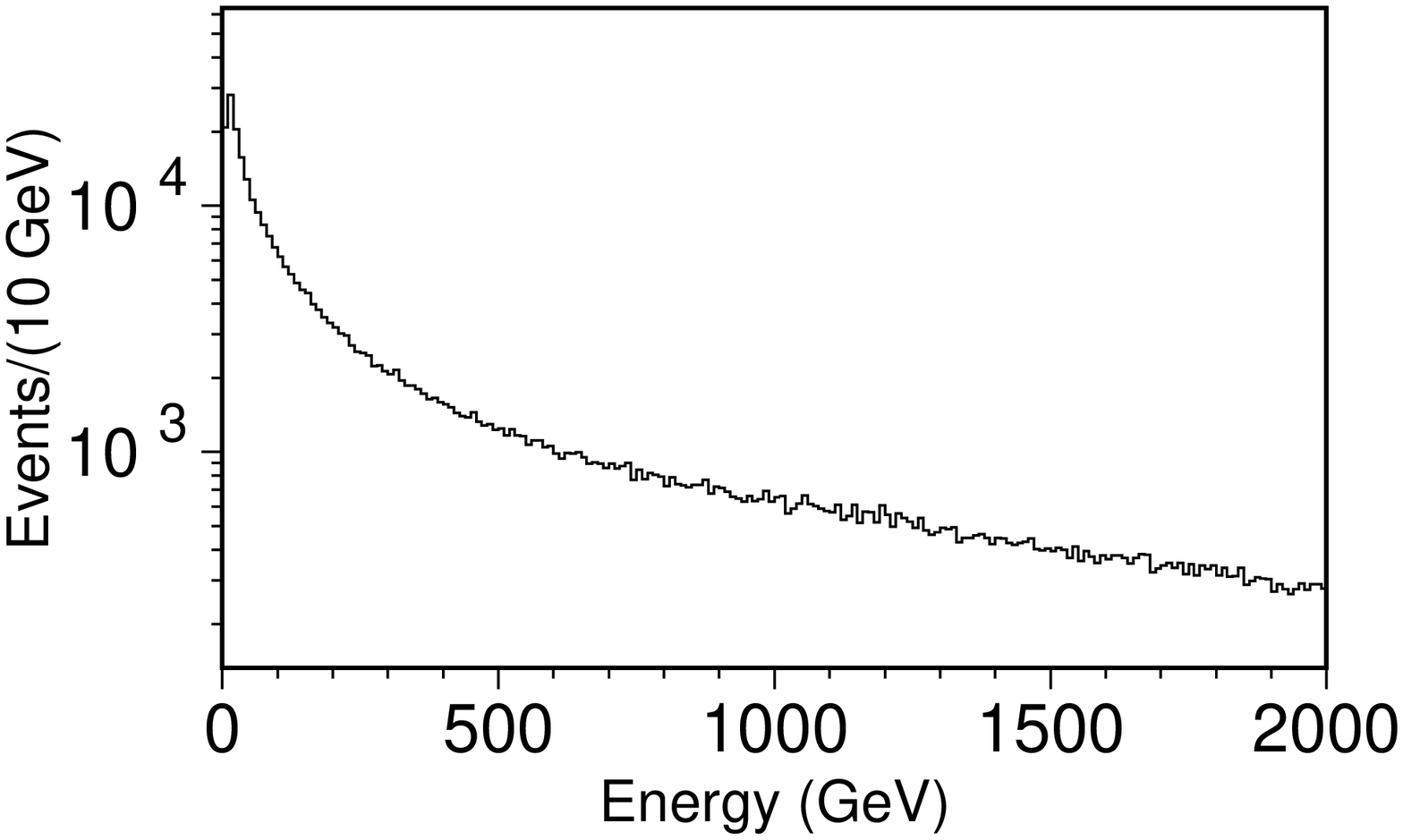,width=10cm}
             \epsfig{file=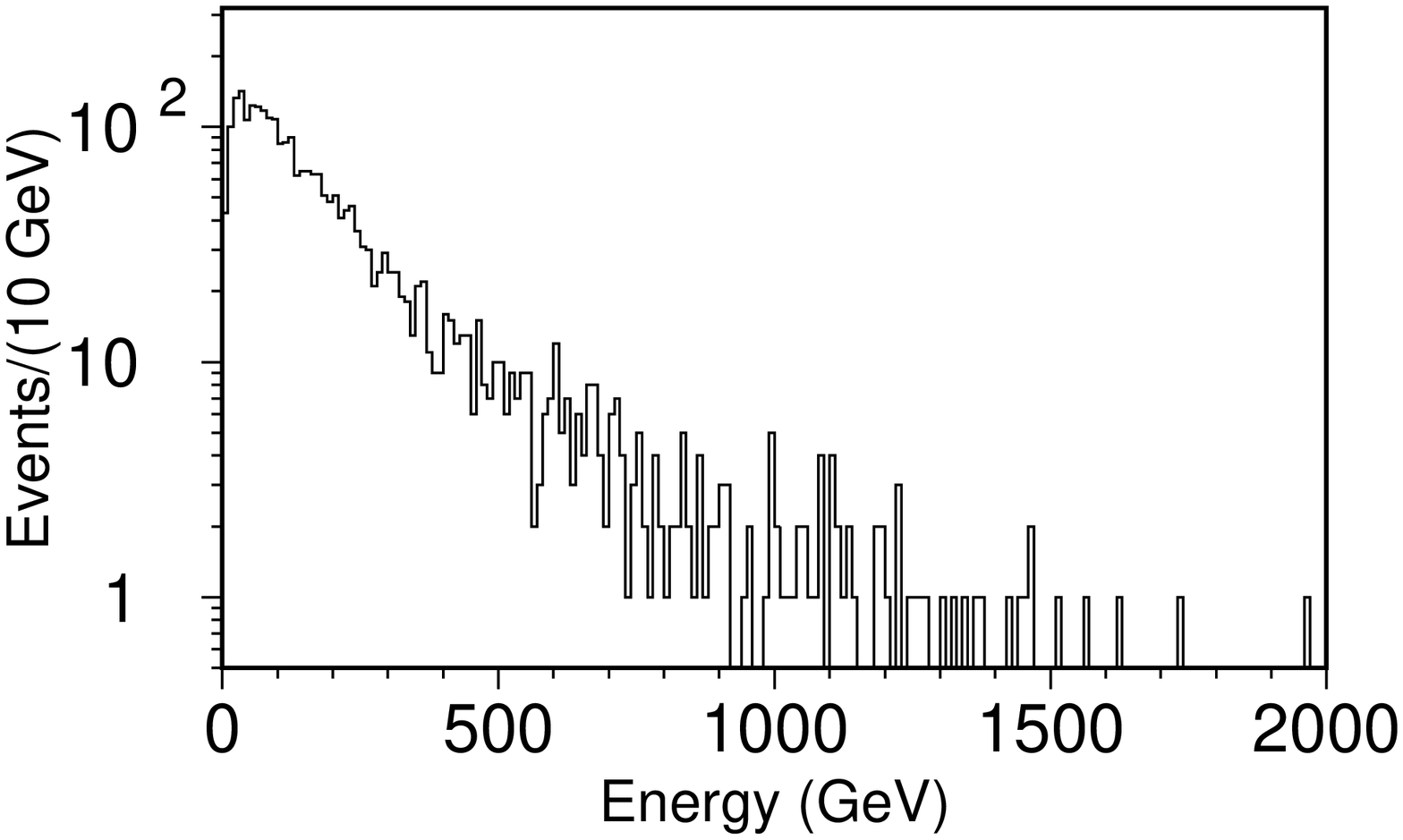,width=10cm}
    \caption{ Energy distributions  for $\nu_e$ (Top), $\nu_\mu$ (Middle) and $\nu_\tau$ (Bottom) with arbitrary scale.}
   \label{fig:energy}}

In order to estimate neutrino fluxes from particle decays produced by pp collisions at $\sqrt{s}=14$~TeV of the LHC, 
we generate a large sample of events with the QCD $2 \to 2$ processes, $q q \to qq$, $q  \bar q \to q  \bar
 q$, $q  \bar q \to g  g$, $q + g \to q g$, $g   g \to  q \bar q $ and $g g \to g  g$, in the PYTHIA.
 In addition, we update  branching fractions for charm and beauty particles reported in the PDG book~\cite{pdg} in the PYTHIA program 
since those heavy particle decays are main sources for $\nu_\tau$, for example, the decay $D_s^+ \to \tau^+ \nu_\tau$, and also
for $\nu_e$ and $\nu_\mu$ production at a few hundred meter away from the interaction point.
The total cross-section for the QCD processes is found to be 54.7 $\hbox{mb}$.  
In particular, the cross sections for charm and beauty particles are
 14.3 $\hbox{mb}$ and 495.7 $\hbox{$\mu$b}$, respectively. 
    
 The produced unstable  particles, even pions and muons, are allowed to decay with their natural life times.
Since it is difficult to estimate yields and directions for   the particles  after they are interacted with beam pipe and detector material,  
 the transverse decay vertexes for neutrinos from the particle decays have to be less than 3 $\cm$ to ensure that 
 the particles are decayed  before hitting the beam pipe.  
 Since two-separated  proton rings at the LHC are merged to the common ring around the interaction point at the LHC, 
 the longitudinal decay vertexes for neutrinos are required to be within 50 $\m$ 
 from the interaction point to guarantee particle decays inside the beam pipe.
 With the requirement of  the transverse vertex, neutrinos are produced by the decays from mainly
 charm and beauty particles.

The average neutrino-beam energy and the number of neutrinos including anti-neutrino per year within 3 $\mrad$ of cone 
in either forward or backward direction
with the nominal luminosity  are listed in Table~\ref{tab:energy}. 
Since neutrinos are associated production with leptons from semileptonic decays of charm and beauty particles and 
the mass of tau lepton is much heavier than electron and muon, 
the energy spectrum and production angle for $\nu_\tau$ are different from ones for $\nu_e$ and $\nu_\mu$.
 Figure~\ref{fig:energy} shows the energy distributions for neutrino and anti-neutrino beams produced within 3 $\mrad$ with respect to proton beam.
The production angle distributions for the neutrino beam are shown in Fig.~\ref{fig:angle}.     

\FIGURE{
               \epsfig{file=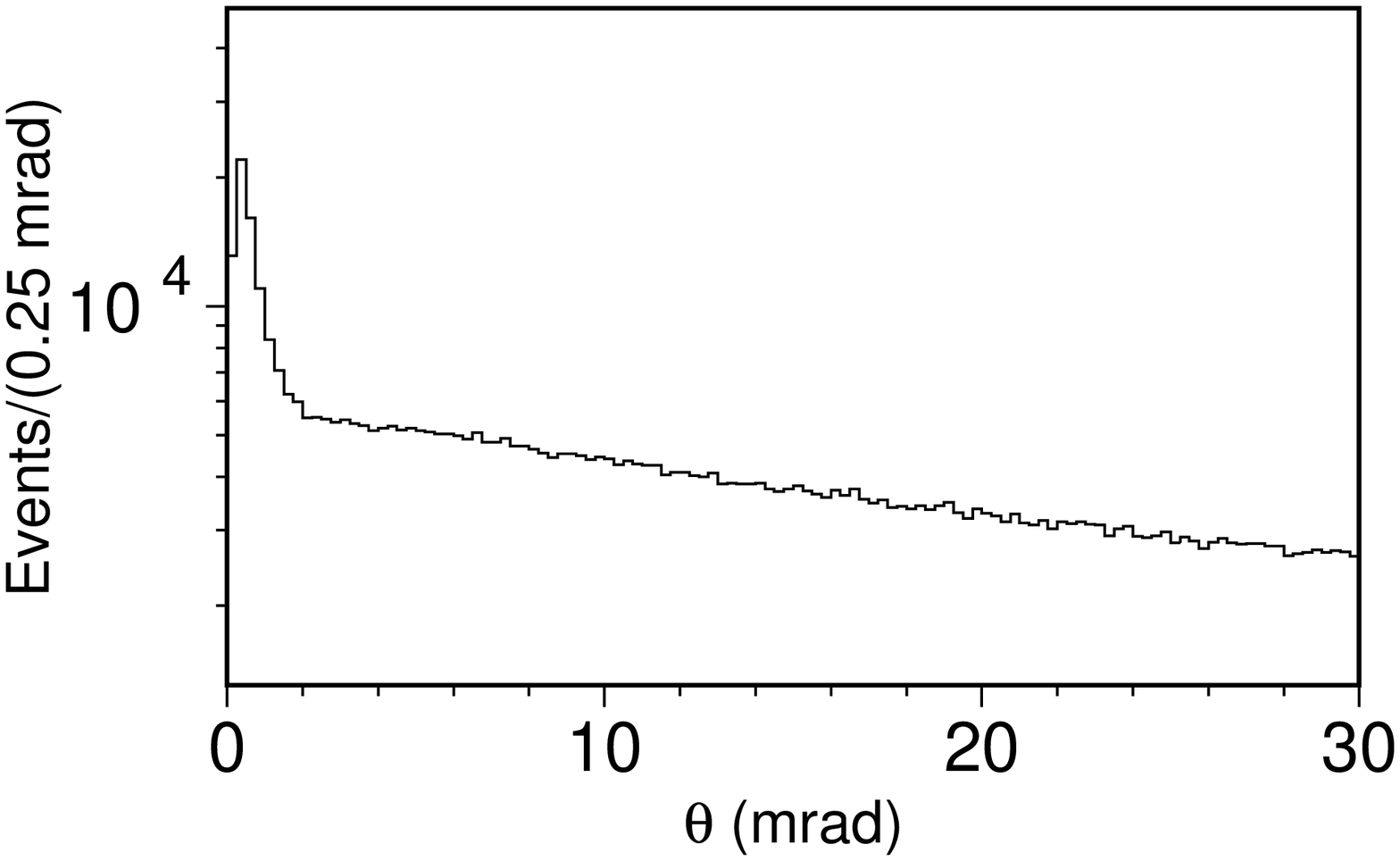,width=10cm}
                \epsfig{file=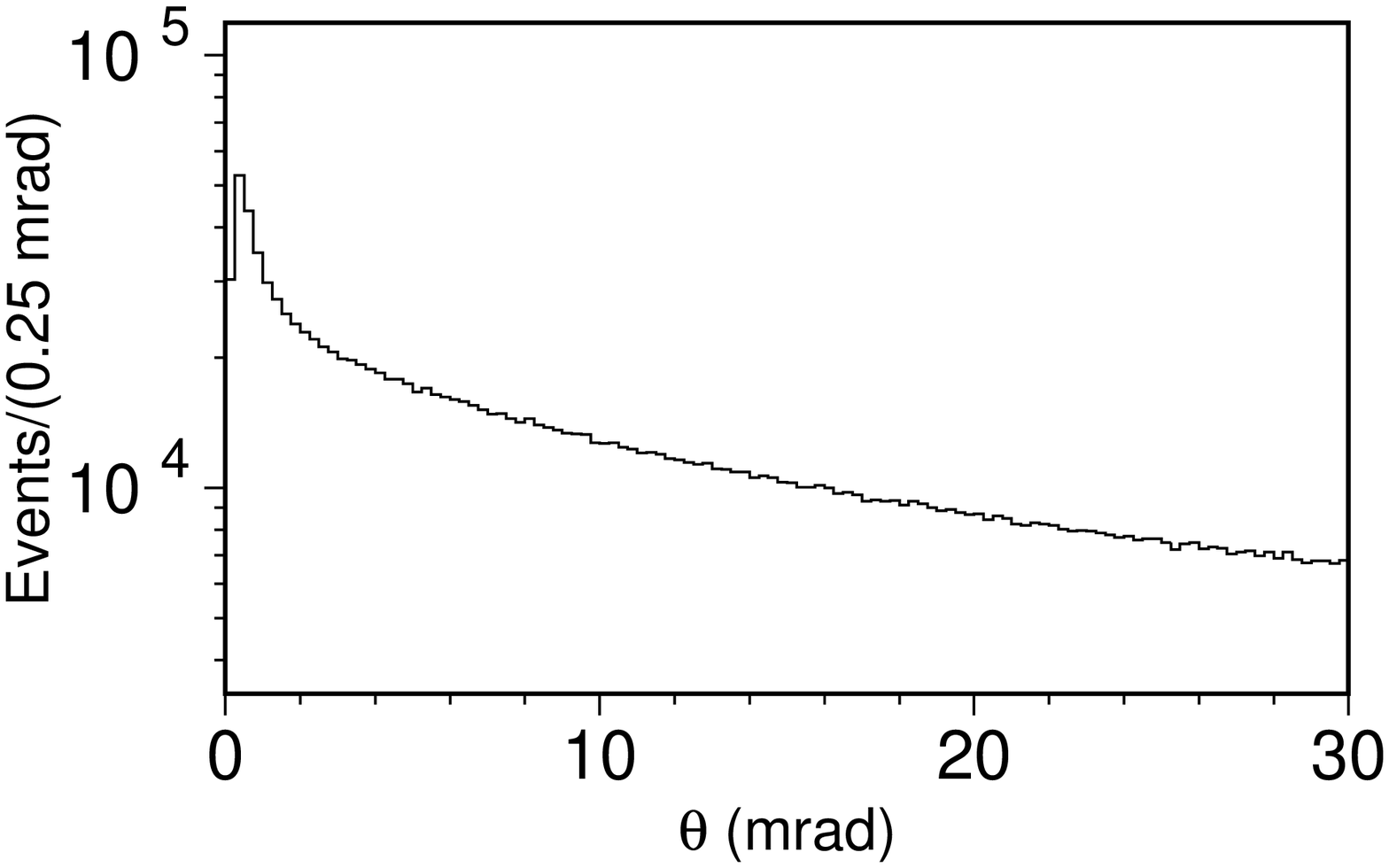,width=10cm}
                \vspace{0.4cm}
             \epsfig{file=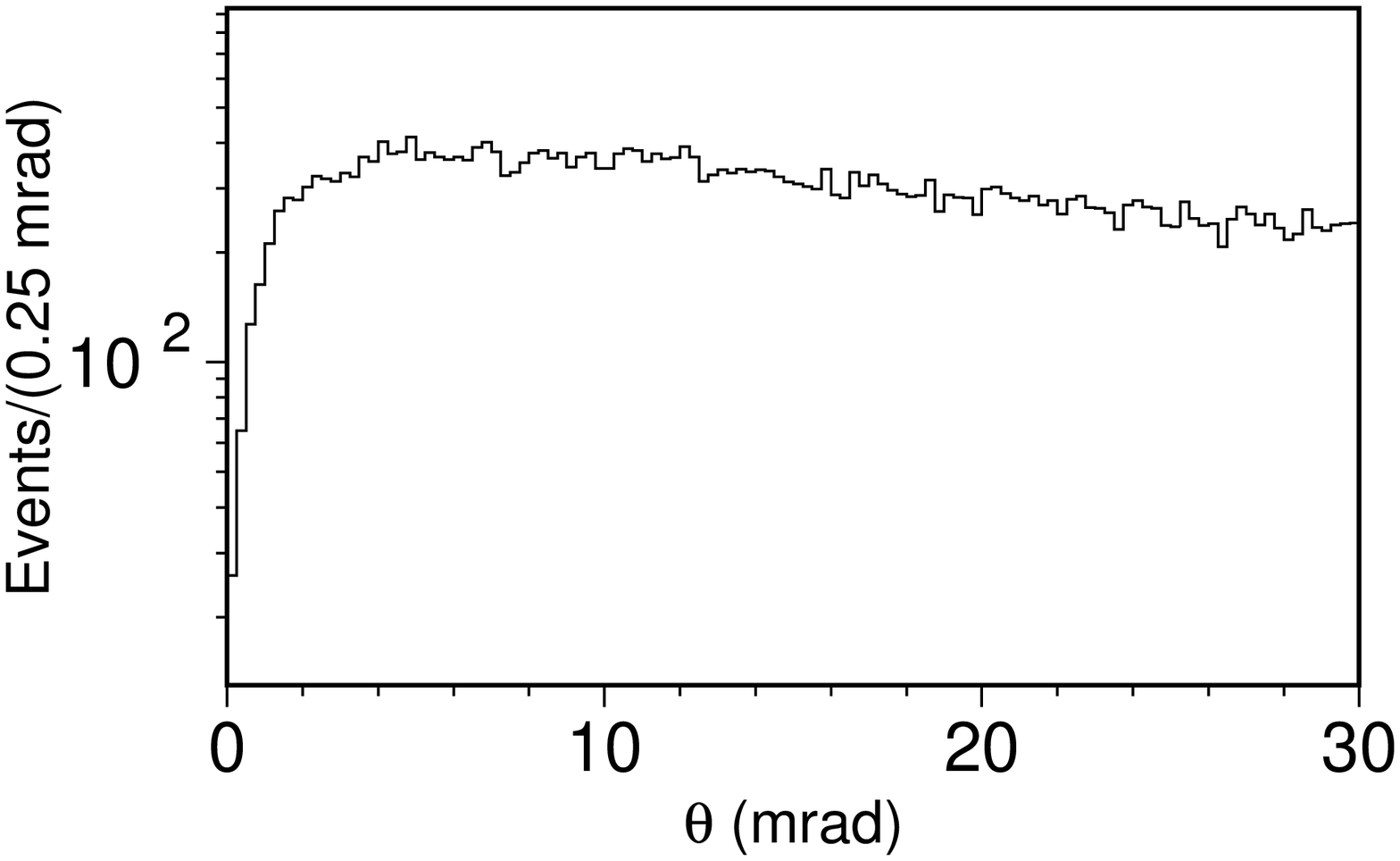,width=10cm}
    \caption{Production angle distributions  for $\nu_e$ (Top), $\nu_\mu$ (Middle) and $\nu_\tau$ (Bottom) with arbitrary scale.}
   \label{fig:angle}}

\section{Proposed Neutrino Experiment at the LHC}

\begin{figure*}[htbp]
\centering
\includegraphics[height=55mm,width=150mm]{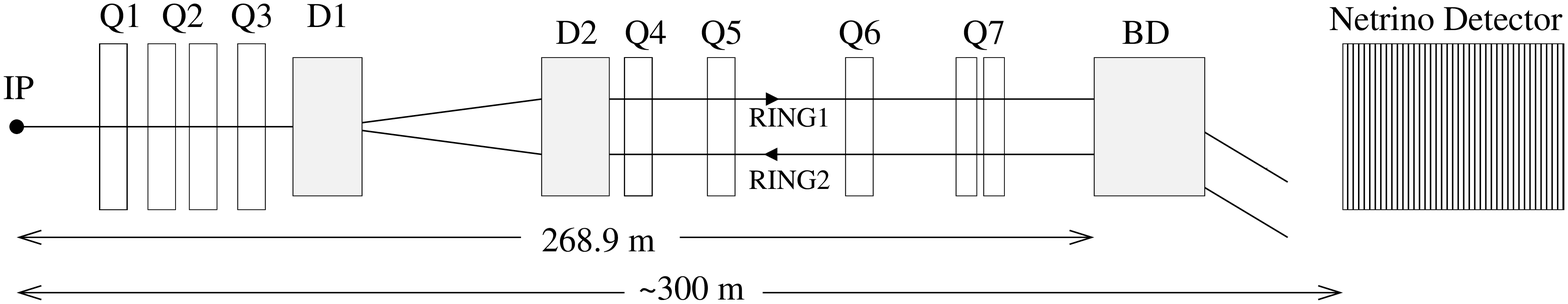}
\caption{ (Not to scale). Layout of a half section of 546-m-long straight section, where IP is the interaction point, 
Q1 to Q7 and D1 to D2 are quadrupole and dipole magnets, respectively, and BD is a bending magnet for matching the arc section to the straight section.
The proton beam in RING1 (RING2) is circulating in clockwise (anticlockwise) direction.}
\label{fig:layout}
\end{figure*}

As the neutrino beam is produced at the interaction point by pp collisions at the LHC, a possible location of neutrino detector is about 300 $\m$
away from the interaction point for high luminosity experiments,  CMS and ATLAS experiments. 
Figure~\ref{fig:layout} shows the layout of  one half of 546-$\m$-long straight section~\cite{lhc} for high luminosity experiments.

In order to detect $\nu_{e}$, $\nu_{\mu}$ and $\nu_{\tau}$ charged current (CC) events with high efficiency, a neutrino detector should have excellent 
capabilities for particle identification and background rejection. A liquid argon time projection chamber~\cite{nudet}, 
which gives high-resolution tracking information and acting as hadronic and electromagnetic calorimeter, would be the best option in current technology.
The CC events for neutrino interacting with detector material are identified by the detection of the charged lepton. 
Identification of $\tau$ leptons from  $\nu_{\tau}$ CC interactions can be performed by detecting  
 a relatively large kink angle and missing transverse momentum 
of   lepton from the decay $\tau^- \to l^- \nu_l \nu_\tau~(l=e,~\mu)$
 compared to $\nu_{e}$ and $\nu_{\mu}$ CC events.
The neutrino and anti-neutrino CC cross sections ($\sigma_{\nu}$)~\cite{pdg} for isoscalar targets are given below,
 \begin{eqnarray}
 \sigma_{\nu}/E_{\nu} = \sigma_0 \times 10^{-38}  \cm^2/\gev, \nonumber
\end{eqnarray} 
where $\sigma_0$ is $0.677 \pm 0.014$ ($0.334 \pm 0.008$) for neutrino (anti-neutrino) and $E_\nu$ is the neutrino energy.

\begin{table}
\centering
 \begin{tabular}{|c|c|c|c|}
\cline{1-4}
  Neutrino              & \multicolumn{3}{|c|}{Expected the number of CC events per year}   \\ \cline{2-4}
  Beam                  & ~~~~~Light & ~~~Heavy & All   \\ \cline{1-4}
  $\nu_e$       &200 (99)   &   31466  (15524)     &   31666 (15623)   \\ \cline{1-4}
  $\nu_\mu$  & 4908 (2421)  &  67990 (33542) &    72898 (37184)  \\ \cline{1-4}
  $\nu_\tau$  &   -  &  209 (103) & 209 (103)  \\ \cline{1-4}
\end{tabular}
\caption{ Summary of the number of neutrino (anti-neutrino) CC events per year with the neutrino detector, 
     liquid argon time projection chamber with 203.3 tons of fiducial-volume, for the nominal LHC luminosity.
     The columns for Light, Heavy and All show 
     the contributions from light, heavy, and light and heavy particle decays, respectively.}
 \label{tab:summ}
 \end{table}

For the estimation of the number of CC events, we assume that the detector is located at 300 m from the interaction point, and the active target for
CC event detection is  subtended 3 $\mrad$ in the forward direction and  50 $\m$ long with a density of 1.36 $\g/\cm^3$ for liquid argon.
The total fiducial volume of this detector is 203.3 tons.

Using the estimated neutrino fluxes per year for $\nu_{e}$, $\nu_{\mu}$ and $\nu_{\tau}$ within 3 mrad in either forward or backward direction
with the nominal LHC luminosity, 
the number of CC events ($N_{CC}$) per year is estimated as follows,
\begin{eqnarray}
 N_{CC} = N_{\nu_l} \sigma_{\nu} N_A \rho L , \nonumber
\end{eqnarray}
where $N_{\nu_l}$ is the neutrino flux for $\nu_l~(l=e,~\mu,~\tau)$, $N_A$ Avogadro's number, $\rho$ the density of detector target material 
and $L$ the length of detector.  In the estimation of $\tau$ neutrino CC events, we require that the energy of $\nu_{\tau}$ has to be larger than
the threshold energy for $\tau$ lepton production,  3.45 $\mgev$.
Table~\ref{tab:summ} is a summary of the number of neutrino and anti-neutrino CC events per year for the neutrino detector described above 
with assumption of 100 \% of detection efficiency.
 
\section{Summary} 
We estimate the neutrino fluxes from proton-proton collisions at  $\sqrt{s}=14$~TeV with the designed luminosity, $10^{34} \lumi$ by using
the PYTHIA program with the tuned parameter set.  
The estimated fluxes for $\nu_e$, $\nu_\mu$ and $\nu_\tau$ including its anti-neutrino from mainly the decays of charm and beauty particles
produced by pp collisions at the LHC
are $2.93 \times 10^{12}$, $7.26 \times 10^{12}$ and $7.1 \times 10^{10}$, respectively. 
For a 203.3 ton detector subtending 3 $\mrad$ at 300 $\m$ away from the interaction point, the number of charged current events for
 $\nu_e$ ($\bar \nu_e$), $\nu_\mu$ ($\bar \nu_\mu$) and $\nu_\tau$ ($\bar \nu_\tau$) are found to be 31,666 (15,623),
 72,898 (37,184) and 209 (103).  
 Our estimated the number of  charged current events for $\nu_\tau$ is lower than one in Ref.~\cite{rujula2} 
 by about a factor of three. 
 This is mainly due to that in our estimation,  the production rate for  $D_s^+$ is  about 0.81 $\hbox{mb}$ while the authors in Ref.~\cite{rujula2} use the production rate for  $D_s^+$ with  2 $\hbox{mb}$.

\acknowledgments
This work was supported in part by the KICOS.

\end{document}